\documentstyle[preprint,prl,aps,epsfig,amssymb,amsmath]{revtex}
\begin{document}
\draft
\title{Diffraction in left-handed materials and theory of Veselago lens}
\author{A. L. Pokrovsky and A. L. Efros}
\address{Department of Physics, University of Utah, Salt Lake City UT, 84112}
\maketitle

\begin{abstract}
A theory of diffraction in the system consisting of  the left-handed and the
right-handed materials is proposed. The theory is based upon the Huygens's principle and 
the Kirchhoff's integral and it is valid if the wavelength is smaller than any relevant
length of the system. The theory is applied to the calculation of the smearing of the
 foci of the Veselago lens due to the finite wavelength. We show  that the Veselago lens
 is a unique optical instrument for the 3D imaging, but it is not a ``superlens'' as it 
has been claimed recently.
\end{abstract}


In his seminal work Veselago \cite{ve} has introduced the concept
of the left-handed materials (LHM's).
In a simplest case the LHM's are materials 
with simultaneously negative
electric permittivity $\epsilon$ and 
magnetic permeability $\mu$ in some frequency range. 
It is easy to show that in the LHM the vectors ${\bf k}, {\bf E}, {\bf H}$
form a left-handed set, while in the usual materials ($\epsilon > 0$, $\mu > 0$)
they form a right-handed set.
If imaginary parts of $\epsilon$ and $\mu$ are small,
the electromagnetic waves (EMW's) propagate in the LHM but they have some 
unusual properties. All these properties originate from the fact that
in the isotropic LHM the Poynting vector 
${\bf S} = {\bf E} \times {\bf H}$ is anti-parallel to the 
wave vector ${\bf k}$.

\begin{figure}
\centering\epsfig{file=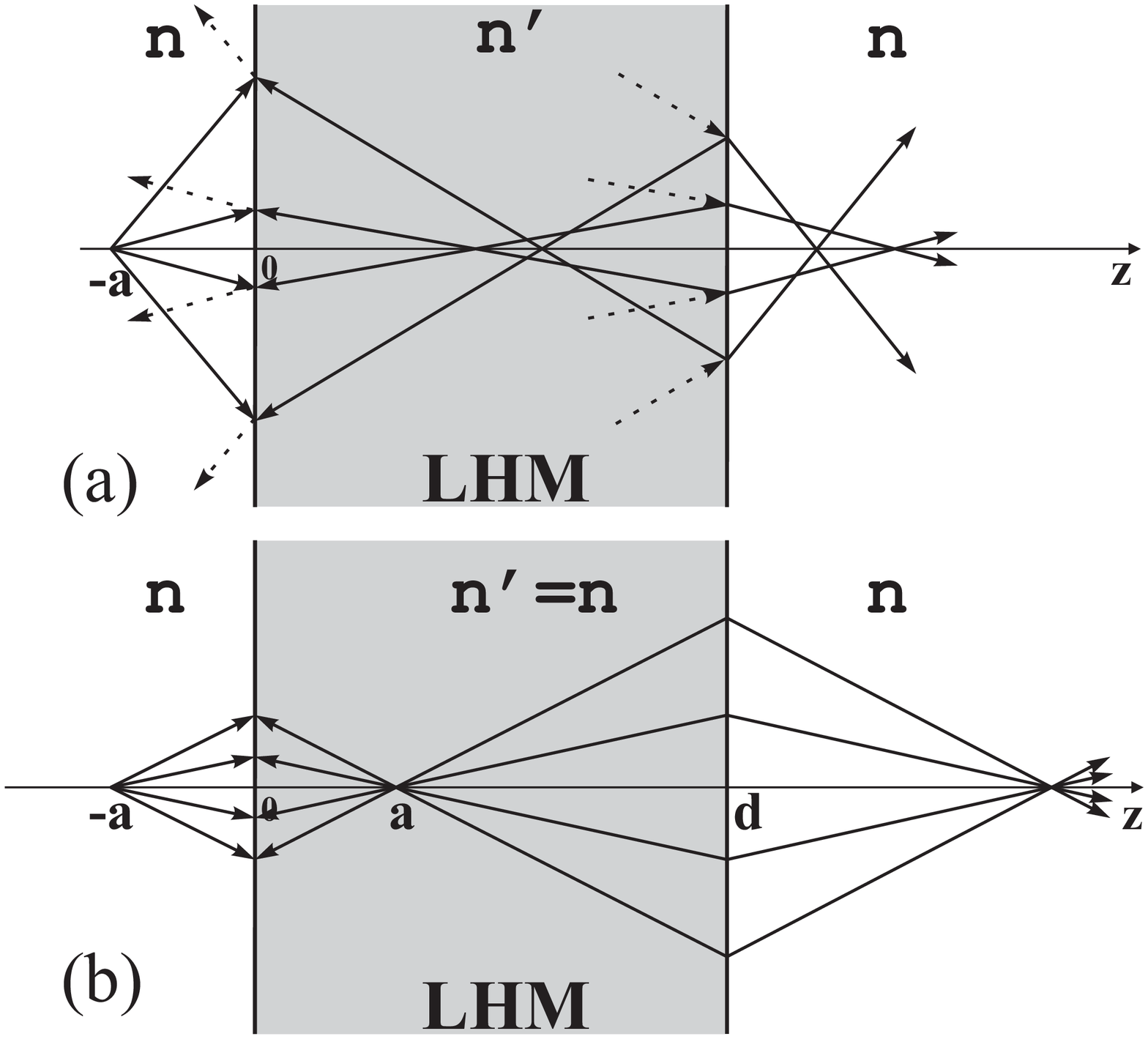, width=8.6cm}
\vspace{0.2cm}
\caption{  Reflection and refraction of light outgoing from a 
point source at $z=-a$ and passing through the slab of the LHM at $0<z<d$.
Refraction of light is described by the anomalous Snell's law.
The arrows represent the direction of the wave vector.
The reflected waves are shown by dashed lines near each interface only.
The slab is surrounded by the usual RHM. (a) $n' > n$. 
(b) The Veselago lens ($n' = n$). The reflected waves are absent, 
all rays pass through two foci.}
\label{fig1}
\end{figure}

Consider a  propagation of the EMW from a point source located 
at the point
$z = -a$ through an infinite slab of the 
LHM with the thickness $d$ and a 
usual right-handed material (RHM) at $z<0$ and $z>d$ (Fig.\ref{fig1}).
It is obvious that $S_z > 0$ everywhere at $z > -a$
because the energy propagates from its source.
The directions of vector ${\bf k}$ for different rays are shown by  arrows. 
They  should be  chosen in such a way that at both
 interfaces tangential components of vector ${\bf k}$
for incident, reflected and refracted waves are the same. Another condition is 
that the component $k_z$ should be parallel to $S_z$ in the RHM and anti-parallel
in the LHM.
Then in the LHM $k_z$ is negative.
It follows that the Snell's law for the RHM-LHM interfaces
has an anomalous form:
$\sin i / \sin r = -n'/n$, where $i$ and $r$ are the angles of incidence and
refraction respectively, $n' = \sqrt{|\epsilon '| |\mu '|/\epsilon_0 \mu_0}$ 
and $n = \sqrt{\epsilon \mu/\epsilon_0 \mu_0}$ are positive refractive indexes
for LHM and RHM respectively. 
The angles of reflection are equal to the angles of incidence.
The  refractive index in the LHM  is often defined  as  negative\cite{sm3},
 but we avoid this definition
 because in an
infinite medium an EMW may propagate in any direction in both LHM and RHM and the 
only physical difference is that vector ${\bf k}$ is directed from the source  of
 the wave in the RHM, while in the LHM it is directed toward the source.

The device shown at Fig.\ref{fig1}(b) is a unique 
optical lens proposed by Veselago. 
In this lens  $\epsilon = -\epsilon '$ and $\mu = -\mu '$, then 
$n' = n$ and $i = -r$.
It is easy to show that at $n = n'$ the reflected wave is completely absent.
Since all the rays going right from the source have $i=-r$,
all of them have foci at points $z=a$ and $z=2d-a$ as shown in Fig.\ref{fig1}(b).
 
All the ideas above have been put forward by Veselago 
about $35$ years ago\cite{ve}.
Recently the method of fabricating of the LHM's on the
basis of metallic photonic crystals has been found and 
the San Diego group has reported the first observation
of the anomalous transmission \cite{sm2} 
and even the anomalous Snell's law \cite{sm3}.
Both observations have been interpreted as the result
of negative $\epsilon$ and $\mu$.
The speculations about the nature of 
negative $\epsilon$ and $\mu$ in photonic crystals
are still controversial (compare \cite{sm3,sm2,p2,so,pok}), 
but the very existence of the LHM seems to be demonstrated.  
Since the LHM's become reality it is time to develop a 
deeper understanding of their electrodynamic properties in order to use
the advantages of these materials.

One can see that the Veselago lens, shown at Fig.\ref{fig1}(b), 
is an {\it absolute instrument} because it images 
stigmatically a three-dimensional domain $-d \le z \le 0$ and the 
optical length of any curve in the object space is equal to the optical length of
its image\cite{born}.
The only other absolute instrument we are aware of is the famous
``fish-eye'' of Maxwell \cite{born}.
Note, that the definition of the absolute instrument assumes 
geometrical optics only.
Since the LHM's have been already obtained  we think that the 
Veselago lens can be extremely important device  for the  
3D imaging.

Pendry \cite{p2} claims that the Veselago lens has a different unique property.
Due to Pendry the resolution of this lens does not have a 
traditional wavelength limitation which follows 
from the uncertainty principle.
Pendry has introduced a new term  ``superlenses'' with the Veselago 
lens as
a first representative of this class. 
Two comments appeared recently \cite{com1,com2}
where the work of Pendry was  criticized.

In this paper we propose a general scalar theory of diffraction in the LHM which
is based upon the Huygens's principle and the Kirchhoff's integral. 
As any diffraction approach our
theory works under  condition that the wavelength is much smaller
than any relevant geometrical length in the problem.
We apply this theory to the Veselago lens and calculate the smearing 
of the foci due to the finite wavelength.
Thus, our result does not support the idea of  ``superlens''.
The discrepancy between our result and previous ones is analyzed.

The first problem is to find the Green function for the Helmholtz 
equation for the LHM which describes propagation of a 
spherical wave from the point source.
It is easy to show that it has a form
$\exp{(-i k R)}/R$, where $k=\omega n/c$ and  $R$ is a distance from the source.
At a small element of the sphere $R = const$ the spherical wave 
can be considered as a plane wave which is characterized by 
the Poynting vector ${\bf S}$ and wave vector ${\bf k}$ both with  the radial
component only.
Since ${\bf S}$ is directed along the external normal to the 
surface element, the wave vector ${\bf k}$ in the LHM is directed 
along the  internal normal.
It is easy to see that our Green function obeys these properties.

Following the principles of the scalar theory of diffraction\cite{lan}
the field $u$ at the observation point $P$ can be written in a form
of a surface integral
\begin{equation}
\label{a}
u_P =  b_l \int u \frac{e^{-i k R}}{R} d f_n,
\end{equation} 
where $R$ is the length of the vector from the point $P$ to the surface element,
 $d f_n$ is the projection of the surface  element $d f$
on the plane perpendicular to the direction of the ray coming 
from the source
to $d f$, $b_l$ is a constant for any LHM.
To find  $b_l$ one can consider a plane wave with the
wave vector normal to the infinite plane of integration. 
Since this plane is 
fictional, the constant can be found from the condition that 
the Huygens's principle in the form Eq.(\ref{a}) 
reproduces the same plane wave.
Doing the  calculations similar to  Ref.\cite{lan} 
one gets $b_l = -k/ 2 \pi i$, so that
for the LHM the constant  $b_l$ has a different sign than 
the similar constant $b_r$ for the RHM.

The Huygens's principle can be applied to any interface which
has a curvature larger than the wavelength.
It gives the correct direction of refracted waves but it does not 
give the amplitudes of both refracted and reflected waves.
However, it can be successfully applied to the Veselago lens 
where reflected waves are absent.

Note, that there are some other methods to describe 
the diffraction which may be also used 
if the source of the rays is unknown.
They are described and compared in details in 
Jackson's textbook \cite{jac}.
One can show that all the methods give the same result at $r=-i$.

\begin{figure}
\centering\epsfig{file=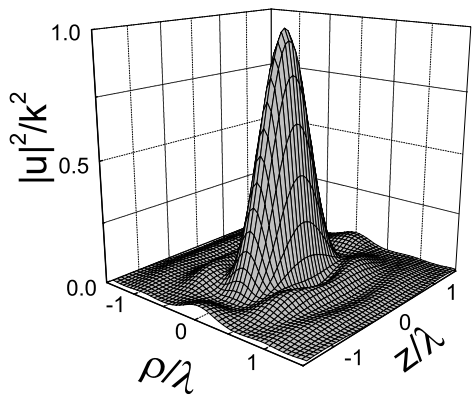, width=8.6cm}
\vspace{0.2cm}
\caption{Distribution of the dimensionless square modulus of the scalar field $|u|^2/k^2$ 
near the foci of the Veselago lens as a function of $\rho$ and $z$ as
given by Eq.(\protect\ref{c}). Here  $\lambda=2 \pi/k$ is the wavelength.}
\label{fig2}
\end{figure}

Now we apply Eq.(\ref{a}) to the Veselago lens.
To find the field $u$ inside the slab we shall 
integrate in Eq.(\ref{a})
over the plane $z=0$. The field $u$ in this plane is
produced by a point source and has a form 
\begin{equation}
u(x, y, 0) = \frac{e^{i k \sqrt{a^2 + x^2 + y^2}}}{\sqrt{a^2 + x^2 + y^2}}.
\end{equation}
The field inside the slab can be found using Eq.(\ref{a}) with a 
constant $b_{rl}$ instead of $b_l$ because now we are 
integrating over the 
 RHM-LHM interface rather than over the fictional surface in the LHM.
In a similar way at the LHM-RHM interface one should use a constant $b_{lr}$.
Using the method described in Ref.\cite{lan}
it is easy to show that $b_{rl} = b_l$ and $b_{lr} = b_r$.
Thus one gets
\begin{equation}
\label{b}
u(x, y, z) = b_l a 
\int\limits_{-\infty}^{\infty} 
\int\limits_{-\infty}^{\infty} 
\frac{e^{i k \sqrt{a^2 + x_1^2 + y_1^2}}}{a^2+x_1^2+y_1^2} 
\frac{e^{-i k \sqrt{z^2 + (x_1-x)^2 + (y_1-y)^2}}}
{\sqrt{z^2+(x_1-x)^2 + (y_1-y)^2}}\: dx_1 dy_1,
\end{equation}
where the additional factor $a/\sqrt{a^2+x_1^2+y_1^2}$ is the cosine of the angle
between the ray, coming from the source to the point $\{x_1, y_1, 0 \}$
and the unit vector in $z$ direction.
One can see that 
the optical lengths for all rays 
(the sum of exponents in the integrand of Eq.(\ref{b})) 
from the point source to the focus, located at
$z = a$, $x = y = 0$, are zero and
the value of the field at the focal point $u(0, 0, a) = i k$, while the geometrical 
optics gives an infinite field in this point.
To find $u$ in the vicinity of the focus  one should
expand the integrand in Eq.(\ref{b}) near the point $(0, 0, a)$
assuming $x \ll a$, $y \ll a$ and $|\zeta| \ll a$, where $\zeta = z-a$. One gets
\begin{equation}
\label{c}
u(\rho, \zeta) = k \left[ \frac{i \sin{(k \sqrt{\rho^2+\zeta^2})}}
{k \sqrt{\rho^2+\zeta^2}} - 
\int\limits_{0}^{1} J_0(k \rho \sqrt{1-s^2}) \sin{(k \zeta s)} ds \right],
\end{equation}
where $\rho^2 = x^2+y^2$.
At $\rho = 0$ one gets analytical expression
\begin{equation}
u(0, \zeta) =  \frac{1-\cos{(k \zeta)} + i \sin{(k \zeta)}}{ \zeta }. 
\end{equation}
Another analytical expression can be obtained at $\zeta = 0$
\begin{equation}
u(\rho, 0) =  \frac{i \sin( k \rho) }{\rho}.
\end{equation}
Figure \ref{fig2} shows dimensionless function $|u(\rho, \zeta)|^2/k^2$ 
as given by Eq.(\ref{c}).
One can see that the smearing of the focus is anisotropic.
The half-width in $z$ direction is approximately one wavelength while
in $\rho$ direction it is approximately twice as less.
At small $x, y, \zeta$ the surfaces of a constant $|u(x,y,\zeta)|^2$ are
ellipsoids of revolution along $z$ axis.

Now we find the field $u$ in the close vicinity of the second focus located
at $x = y = 0$, $z = 2d-a$.
The general expression for $u$ at  $z>d$ 
differs from Eq.(\ref{b}).
One should apply the Huygens's principle  to both interfaces located at  $z=0$ and  $z=d$.
The later one is the  
 LHM-RHM interface and 
the constant  $b_r = k/2 \pi i$ should be used instead of $b_l$.
One gets an additional integral over the plane $z=d$ so that expression 
for the field has a form
\begin{multline}
\label{sf0}
u(x, y, z) = b_l b_r a d 
\int\limits_{-\infty}^{\infty}  \int\limits_{-\infty}^{\infty} \: dx_1 dy_1 
\int\limits_{-\infty}^{\infty}  \int\limits_{-\infty}^{\infty} \: dx_2 dy_2 
\frac{e^{i k \sqrt{a^2 + x_1^2 + y_1^2}}}{a^2+x_1^2+y_1^2}  \\
\frac{e^{-i k \sqrt{d^2 + (x_1-x_2)^2 + (y_1-y_2)^2}}}{d^2+(x_1-x_2)^2+(y_1-y_2)^2} 
\frac{e^{-i k \sqrt{(z-d)^2 + (x_2-x)^2 + (y_2-y)^2}}}{\sqrt{(z-d)^2+(x_2-x)^2+(y_2-y)^2}}.
\end{multline}
To calculate these integrals in the vicinity of the second focus
using inequalities $k d \gg 1$, $ka\gg 1$ one should introduce new variables $\{ s,t \}$
instead of variables $\{x_2, y_2\}$ by relations 
\begin{gather}
\label{lines}
x_2 = -(\frac{d}{a} - 1) x_1 + s \\
y_2 = -(\frac{d}{a} - 1) y_1 + t.
\end{gather}
Equation (\ref{lines}) has the following meaning  at $s = t = 0$.
For every point $\{ x_1, y_1\}$ at the first interface
they give a point $\{ x_2, y_2\}$ at the second
interface which is on the ray coming from $\{ x_1, y_1\}$
and passing through the first focus.
Thus the new variables $\{ s,t \}$ describe deviation from the geometrical optics
and they should be small.
One can see from Eq.(\ref{sf0}) that at $s = t = 0$ optical lengths 
of all rays exiting from the point source at $z=-a$
and coming to the second focus at $z=2d-a$ are equal to zero. 
Introducing the new variables and expanding the exponents in Eq.(\ref{sf0})
one can get an expression for the field $u$ in the vicinity of the second focus.
For $\eta = z-2d+a$, $|\eta| \ll a$ one gets
\begin{equation}
\label{ff} 
u(x, y, \eta) = -u(x, y, \zeta)^*|_{\zeta=\eta},
\end{equation}
where the function $u(x, y, \zeta)$ is given by Eq.(\ref{b}).
In other words, the smearing of the $|u(x, y, \eta)|^2$ in the second focus 
is the same as the smearing in the first one.

Equation (\ref{ff}) can be also obtained in a more physical way.
One can calculate field $u$ far from the foci expanding integrand 
in Eqs.(\ref{b},\ref{sf0}) near the geometrical rays as it was described above.
In this case the result is  exactly the same as in the geometrical optics.
Namely, in the region $0<z<d$ 
\begin{equation}
\label{far}
u(x, y, z) = \frac{e^{i k (R - R_1)}}{R - R_1},
\end{equation}
where $R = a \sqrt{1+(x^2 + y^2)/(a-z)^2}$, $R_1 = z \sqrt{1+(x^2 + y^2)/(a-z)^2}$.
At $z>d$ one should substitute into Eq.(\ref{far}) 
$R = (z-d) \sqrt{1+(x^2 + y^2)/(2 d-a-z)^2}$, $R_1 = 
(d-a) \sqrt{1+(x^2 + y^2)/(2 d-a-z)^2}$. The expression in Eq.(\ref{far}) becomes 
infinite in both foci as it should be in the framework of the geometrical optics. However,
it can be used to  calculate field $u$ at the plane $z=d$
\begin{equation}
\label{n}
u(x, y, d) = -\frac{e^{-i k\sqrt{(d-a)^2 + x^2 + y^2} }}{\sqrt{(d-a)^2 + x^2 + y^2}}.
\end{equation}
Note that the minus sign in Eq.(\ref{n}) results from the passing of the rays through the
first focus. This gives extra phase $\pi$.

Now we can forget about the region $z<d$ and 
apply the Huygens's  principle to the $z=d$ to find the field near the second focus.
The field will be described by Eq.(\ref{a}) with the positive exponent
and $b_r=k/2 \pi i$. Finally we get the result which is connected with 
Eq.(\ref{b}) by Eq.(\ref{ff}).

Using Eq.(\ref{far}) one can calculate the flux of 
energy through any plane
perpendicular to $z$ axis for $z>0$.
One can show that it is independent on $z$ and equal to $2 \pi$ 
in our units. 
Note, that the flux of energy through the hemisphere around 
the point source
at $z=-a$ defined as $\int |u|^2 d f_n$ is equal  2$\pi$, since
$u=\exp{(i k R)}/R$.

Now we compare our results with the analytical calculations 
of Pendry \cite{p2} and Ziolkowski and Heyman \cite{ziol}.
Both papers claim that the Veselago lens in the ideal 
(lossless) regime is a ``superlens'', which is able to provide
a perfect focusing. 
In both papers the spherical wave outgoing from the source
is represented as a superposition of plane waves,
which are fictitious and do not correspond to
the cylindrical symmetry of the problem.
This superposition contains the ``evanescent'' waves (EW's), for which
$k_x^2 + k_y^2 > \omega^2/c^2$.
One can easily show that the Poynting vector of each EW
has a non zero components in the $x$-$y$ plane but zero
component in $z$ direction.
It follows from the second observation that 
the contribution of the EW's to the intensity near 
the foci of the Veselago 
lens should be exponentially small if $a \omega/c \gg 1$.
Pendry has explained perfect focusing as a result of 
amplification of EW's by the LHM.

One can see from Eq.(47a) of Ref.\cite{ziol} that the 
amplitude of a single EW increases exponentially in the LHM
with increasing distance $z$ from the source.
Since the LHM is a passive medium, we think, that these
EW-solutions should be omitted as nonphysical.
The mathematical inconsistency of these solutions
can be seen from the fact that 
the integral (Eq.(38) of Ref.\cite{ziol}), 
that describes the superposition of the 
plane waves, diverges at large $k_x^2 + k_y^2$ 
in the interval $a < z <2d-a$ at any value of $x$ and $y$.
Note, that the contribution of propagating waves into this integral
is finite and it coincides with our result near the foci.

The advantage of the diffraction theory is 
that it is a regular
perturbation with respect to $1/k d$. We think that 
the EW's never appear
in this theory because their contribution is of the order of 
$\exp{(-2 k d)}$. 

The computations, performed in Ref.\cite{ziol},
do not show any focusing.
We think that the main reason is that
their $a$ and $d$ are of the order of the wavelength.
Our calculations do not predict any focusing for such wavelength.

Finally, we have proposed the theory of diffraction in a system, 
consisting of the LHM and the RHM and have applied this  theory
to the calculation of the smearing of the foci of the Veselago lens. 
This smearing is of the order of the wavelength so, 
from this point of view, the Veselago lens does not differ
from any other lens.

The work has been funded by the NSF grant DMR-0102964.


\end{document}